\documentclass[RNAAS]{aastex62}
\usepackage{amsmath}
\newcommand{\tp}{t_{\rm{peak}}}
\newcommand{\Lp}{L_{\rm{peak}}}

\begin{document}

\title{Analytical calculation of the numerical results of Khatami and Kasen for transient peak time and luminosity}

%% Note that the corresponding author command and emails has to come
%% before everything else. Also place all the emails in the \email
%% command instead of using multiple \email calls.
\correspondingauthor{Doron Kushnir}
\email{doron.kushnir@weizmann.ac.il}

\author{Doron Kushnir}
\affiliation{Dept. of Particle Phys. \& Astrophys., Weizmann Institute of
Science, Rehovot 76100, Israel}

\author{Boaz Katz}
\affiliation{Dept. of Particle Phys. \& Astrophys., Weizmann Institute of
Science, Rehovot 76100, Israel}

%% Note that RNAAS manuscripts DO NOT have abstracts.
%\keywords{editorials, notices --- 
%miscellaneous --- catalogs --- surveys}

%% Start the main body of the article. If no sections in the 
%% research note leave the \section call blank to make the title.
\section{} 

The diffusion approximation is often used to study supernovae light-curves around peak light, where it is applicable \citep[e.g.][]{Arnett1982,Pinto2000,Khatami2019}. By analytic arguments and numerical studies of toy models, \citet{Khatami2019} recently argued for a new approximate relation between peak bolometric Luminosity, $\Lp$, and the time of peak since explosion, $\tp$, for transients involving homologous expansion:  
\begin{equation}\label{eq:beta def}
\Lp=\frac{2}{(\beta\tp)^2}\int_0^{\beta\tp}t'Q(t')dt',
\end{equation}
where $Q(t)$ is the heating rate of the ejecta, and $\beta$ is an order unity parameter that is calibrated from numerical calculations. \citet{Khatami2019} demonstrated its validity using Monte-Carlo radiation transfer simulations of ejecta with homogenous density and (for most cases considered) constant opacity. Interestingly, constant values of $\beta$ accurately reproduce the numerical calculations for different heating distributions and over a wide range of energy release times. Here we show that the diffusion and the adiabatic loss of energy in homologous expansion is equivalent %(under a suitable change of variables) 
to a static diffusion equation and provide an analytic solution for the case of uniform density and opacity \citep[extending the results of][]{Pinto2000}. Our accurate analytical solutions reproduce and extend the results of \citet{Khatami2019} for this case, allowing clarification for the universality of Eq. \eqref{eq:beta def} as well as new limitations to its use. 

Assuming non-relativistic homologous expansion, with radiation dominated pressure, the diffusion of bolometric radiation energy is given by \citep[e.g., eq. 10 of][]{Pinto2000}
%\citep[e.g., eq. 10 of][with $E$ replaced here with $e$ and $R/\dot R$ with t]{Pinto2000}
\begin{equation}
\frac{De}{Dt}+\mathbf{\nabla}\cdot\left(\frac{c}{3\kappa\rho}\mathbf{\nabla}e\right)+4\frac{e}{t}=\epsilon,
\end{equation}
where $e$ is the energy density in radiation, $\kappa$ is the effective opacity (which may vary in time and position), $\rho$ the density and $\epsilon$ the local energy generation rate of radiation per unit volume. Working in velocity coordinates $\mathbf{v}=\mathbf{x}/t$ (with spatial derivatives related by $\nabla_v=t\nabla$) and using the scaled quantities \citep[similar to][]{Arnett1982}:
\begin{equation}
\tilde e= t^4e,~~\tilde\rho =t^3\rho,~~\tilde\epsilon=t^3\epsilon,~~\tilde t = \frac12t^2 \implies d\tilde t=tdt,
\end{equation}
a static diffusion equation is obtained:
\begin{equation}
\frac{\partial \tilde e}{\partial \tilde t}+\mathbf{\nabla_v}\cdot\left(\frac{c}{3\kappa\tilde\rho}\mathbf{\nabla_v}\tilde e\right)=\tilde \epsilon.
\end{equation}
Note that $\tilde\rho(\mathbf{v})$ is independent of time and that the opacity does not require scaling (though it is usually time and space dependent). The scaled global energy generation rate $\tilde Q$, luminosity $\tilde L$ and total energy in radiation $\tilde E$ are related to their physical counterparts $Q,L$ and $E$ by:
\begin{equation}
\tilde Q=  Q,~~\tilde L= L,~\tilde E = Et.
\end{equation}
Total energy conservation reads:
\begin{equation}
\tilde E=\tilde E|_{\tilde t=0}+\int \tilde Q d\tilde t - \int \tilde L d\tilde t  ~~~\Leftrightarrow~~~  Et=(Et)|_{t=0}+\int Q tdt - \int L tdt,
\end{equation}
where the rhs is correct regardless of the diffusion approximation \citep{Katz2013}. 
Eq. \eqref{eq:beta def} is expressed in the new variables as
\begin{equation}\label{eq:beta def static}
 L_{\rm peak}= \frac{\int_0^{\tilde t_{\beta}} Q(\tilde t') d\tilde t'}{\tilde t_{\beta}}\equiv\langle Q \rangle_{\tilde t_{\beta}},\end{equation}
where $\tilde t_{\beta}=\beta^2\tilde t_{\rm peak}$, and can be stated as $ L_{\rm peak}$ being equal to the average deposition $Q$ from $0$ to $\tilde t_{\beta}$.

Following \citet{Khatami2019}, we consider depositions that have a constant spatial distribution and a magnitude that is decreasing over time with a typical decay time of $t_s$ (specifically, $Q(t)=Q_0e^{-t/t_s}$). The luminosity can therefore be expressed as:
\begin{equation}\label{eq:L of Ldelta}
L(\tilde t)=\int_0^{\tilde t} d\tilde t' Q(\tilde t') L_{\delta}(\tilde t - \tilde t')
\end{equation}
where $L_{\delta}(\tilde t)$ is the luminosity obtained in the impulse approximation, $Q_{\delta}=\delta(\tilde t)$, for the same spatial distribution of the deposition.

It is useful to consider Eq. \eqref{eq:beta def static} in two extremes. First, for a deposition time $t_s$ which is much longer than the diffusion time $t_d$, $\tilde Q$ is approximately constant beyond $t_{\rm peak}$ and equal to $\tilde L_{\rm peak}$. To see this, note that for constant deposition, $\tilde L$ is a monotonically increasing function (accumulation of $L_{\delta}$, Eq~\eqref{eq:L of Ldelta}) that is approximately equal to $\tilde Q$ at times much longer than the diffusion time.  In this case, Eq.~\eqref{eq:beta def static} is correct \emph{for any value} of $\beta$ of order unity.  Second, for a deposition time which is much shorter than the diffusion time (\textit{the impulse limit}), the integral on the rhs of Eq.~\eqref{eq:beta def static} reduces to the total deposition energy $\tilde E$, which is independent of $\beta$ and the specific source function. Therefore, Eq.~\eqref{eq:beta def static} becomes correct for the choice $\beta_{\rm{imp}}=(\tilde E/ (L_{\rm peak}\tilde t_{\rm peak}))^{1/2}$.  These arguments imply that Eq.~\eqref{eq:beta def static} (and thus Eq.~\eqref{eq:beta def}) is correct for $\beta=\beta_{\rm{imp}}$ for both very short $t_s\ll t_d$ and very long $t_s\gg t_d$ deposition times. 
%But what about intermediate values of $t_s\sim t_d$?
\citet{Khatami2019} argued that for the special case of a uniform ejecta (uniform density and opacity), a single value of $\beta$ applies to a good approximation also for intermediate values of $t_s\sim t_d$. We show below that while this is true for central deposition, this is not the case for the extended depositions considered by \citet{Khatami2019}.
%It was shown by  \citet{Khatami2019} that for the special case of a central deposition in a uniform ejecta (uniform density and opacity), the same value of $\beta\approx 4/3$ applies to a good approximation at all values of $t_s$. We show below that this is not the case however for the extended depositions considered by \citet{Khatami2019}.

\begin{figure}[h!]
\begin{center}
\includegraphics[scale=0.8,angle=0]{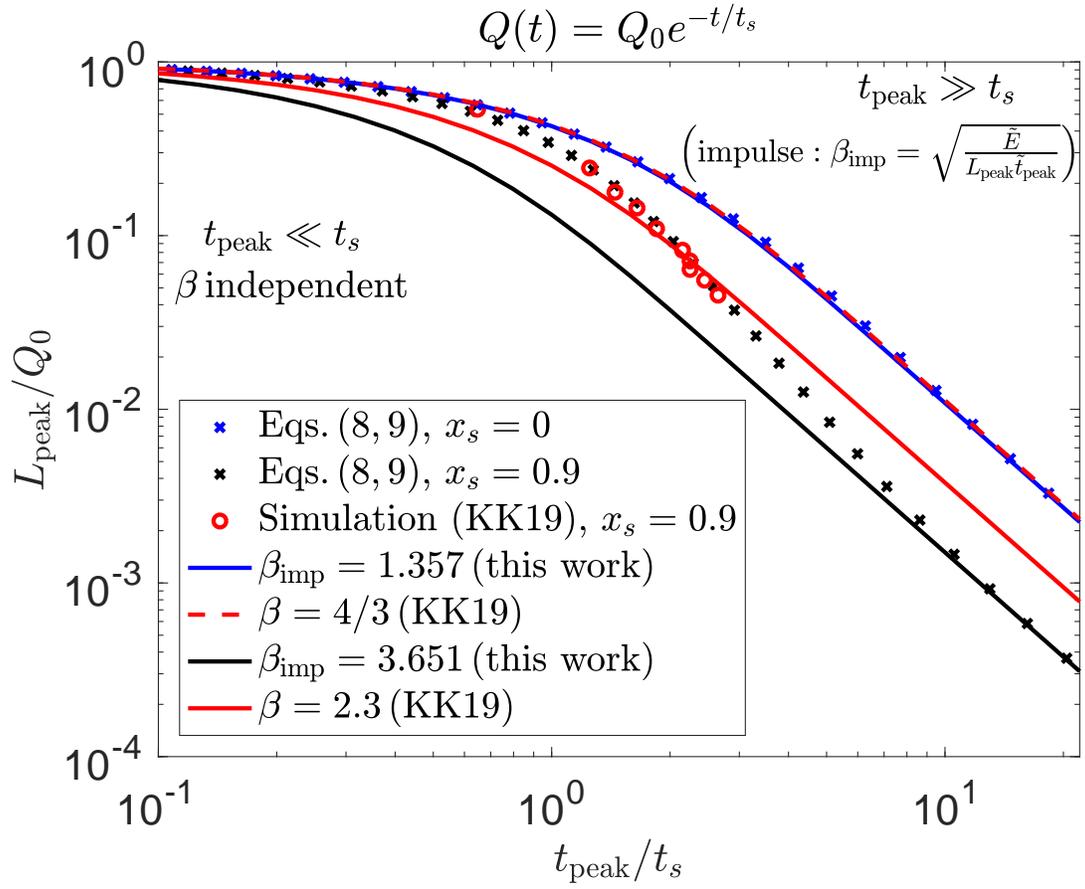}
\caption{Peak times and luminosities for the ejecta discussed in the text. The analytical results (equations~Eqs.~(\ref{eq:L of Ldelta}-\ref{eq:Ldelta}), blue crosses, $x_s=0$ and black crosses, $x_s=0.9$) agree and extend the results of the simulations of \citet{Khatami2019} (their figure 8, red dots, shown only for $x_s=0.9$). Solid lines are the analytic solutions of Eq.~\eqref{eq:beta def}, for an exponential deposition \citep{Khatami2019} with corresponding values of $\beta$ as given in the legend. The data behind the figure is provided in
\href{https://webhome.weizmann.ac.il/home/boazkat/AnalyticalHomogenousDiffusion/FigureValues.m}{this link}.\label{fig:1}}
\end{center}
\end{figure}

The luminosity $L_{\delta}$ as a function of time $\tilde t$ from an impulse of energy $\tilde E=1$, deposited uniformly over $x_s$ within a ball of radius $R=1$, and diffusion coefficient $D=1$, is straight forward to derive and is given by
\begin{equation}\label{eq:Ldelta}
L_{\delta}=\frac{6}{x_s^3}\sum_{n=1}^{\infty}(-1)^{n+1}\left(\frac{\sin(n \pi x_s)}{n\pi}-x_s\cos(n\pi x_s)\right)e^{-n^2\pi^2\tilde t}.
\end{equation}
For deposition radii of $x_s=0,0.33,0.9$ the peak times are $\tilde t_{peak}=0.09175,0.08012,0.004500$, the maximum luminosities are $L_{\rm peak}=5.922,5.963, 16.67$ and the values of $\beta_{\rm{imp}}=1/(L_{\rm peak}\tilde t_{peak})^{1/2}$ are $1.357,1.447,3.651$, respectively. While the first two values are similar to the value $\beta=4/3$ reported by \citet{Khatami2019} for small $x_s$, the last value is significantly different from the value of $\beta=2.3$ that they report for the corresponding $x_s=0.9$. The source of the difference can be seen in figure \ref{fig:1},where 
the resulting peak luminosities %for an exponential deposition function $Q=Q_0e^{-t/t_s}=Q_0e^{-(\tilde t/\tilde t_s)^{1/2}}$ 
are shown. For $x_s=0$, a single value $\beta=\beta_{\rm{imp}}\approx4/3$ is a good approximation at all values of $t_s$, while for $x_s=0.9$, the value of $\beta=\beta_{\rm{imp}}=3.651$ fails in the intermediate regime of $t_s\sim t_{\rm peak}$. The value $\beta=2.3$, which \citet{Khatami2019} calibrated to the intermediate regime, does better but fails for very fast deposition. 

A matlab function that calculates the luminosity as a function of time for a given mass, outer velocity, opacity, $x_s$, and $t_s$ using Eqs.~\eqref{eq:L of Ldelta},\eqref{eq:Ldelta}, is provided in \href{https://webhome.weizmann.ac.il/home/boazkat/AnalyticalHomogenousDiffusion/L_of_M_vmax_kappa_xs_ts.m}{this link}.

%% An example table using AASTeX's deluxetable. Note that since
%% only one figure OR one table is allowed this is commented out.
%\begin{deluxetable}{ccl}
%\tablecaption{Example table some English and Greek letters\label{tab:1}}
%\tablehead{
%\colhead{Index number} & \colhead{English} & \colhead{Greek}
%}
%\startdata
%1 & a & alpha ($\alpha$) \\
%2 & b & beta ($\beta$) \\
%3 & c & gamma ($\gamma$) \\
%4 & d & delta ($\delta$) \\
%5 & e & epsilon ($\epsilon$) \\
%\enddata
%\tablecomments{Long tables should only show a short example with the full
%version as a machine readable table with the article.}
%\end{deluxetable}  

\acknowledgments

DK is supported by the Pazi Foundation. BK is supported by the Beracha Foundation and the MINERVA Stiftung.


\begin{thebibliography}{}

\bibitem[Arnett(1982)]{Arnett1982} Arnett, W.~D.\ 1982, \apj, 253, 785 %https://ui.adsabs.harvard.edu/abs/1982ApJ...253..785A

\bibitem[Katz et al.(2013)]{Katz2013} Katz, B., Kushnir, D., \& Dong, S.\ 2013, arXiv e-prints, arXiv:1301.6766

\bibitem[Khatami \& Kasen(2019)]{Khatami2019} Khatami D.~K., Kasen D.~N., 2019, ApJ, 878, 56

\bibitem[Pinto \& Eastman(2000)]{Pinto2000} Pinto, P.~A., \& Eastman, R.~G.\ 2000, \apj, 530, 744

\end{thebibliography}
\end{document}